%% file: main.tex
\documentclass[sigconf]{acmart}

\usepackage{subfigure}
\usepackage{enumitem}

\AtBeginDocument{%
  \providecommand\BibTeX{{%
    Bib\TeX}}}

\acmISBN{978-1-4503-XXXX-X/18/06}

\begin{document}

\title[]{
Navigating User Experience of ChatGPT-based Conversational Recommender Systems: The Effects of Prompt Guidance and Recommendation Domain
}

\author{Yizhe Zhang}
\email{csyzzhang@comp.hkbu.edu.hk}
\affiliation{%
  \institution{Department of Computer Science, Hong Kong Baptist University}
  \streetaddress{}
  \city{Hong Kong}
  \country{China}
}

\author{Yucheng Jin}
\email{yucheng.jin@dukekunshan.edu.cn}
\affiliation{%
  \institution{Duke Kunshan University}
  \streetaddress{}
  \city{Kunshan}
  \country{China}
}
\affiliation{%
  \institution{Hong Kong Baptist University}
  \streetaddress{}
  \city{Hong Kong}
  \country{China}
}

\author{Li Chen}
\email{lichen@comp.hkbu.edu.hk}
\affiliation{%
 \institution{Department of Computer Science, Hong Kong Baptist University}
 \streetaddress{}
  \city{Hong Kong}
  \country{China}
}

\author{Ting Yang}
\email{cstyang@comp.hkbu.edu.hk}
\affiliation{%
  \institution{Department of Computer Science, Hong Kong Baptist University}
  \city{Hong Kong}
  \country{China}
}

\renewcommand{\shortauthors}{}

\begin{abstract}

Conversational recommender systems (CRS) enable users to articulate their preferences and provide feedback through natural language. With the advent of large language models (LLMs), the potential to enhance user engagement with CRS and augment the recommendation process with LLM-generated content has received increasing attention. 
However, the efficacy of LLM-powered CRS is contingent upon the use of prompts, and the subjective perception of recommendation quality can differ across various recommendation domains.
Therefore, we have developed a ChatGPT-based CRS to investigate the impact of these two factors, \textit{prompt guidance} (PG) and \textit{recommendation domain} (RD), on the overall user experience of the system. 
We conducted an online empirical study (N = 100) by employing a mixed-method approach that utilized a between-subjects design for the variable of PG (with vs. without) and a within-subjects design for RD (book recommendations vs. job recommendations).
The findings reveal that PG can substantially enhance the system's explainability, adaptability, perceived ease of use, and transparency. Moreover, users are inclined to perceive a greater sense of novelty and demonstrate a higher propensity to engage with and try recommended items in the context of book recommendations as opposed to job recommendations. Furthermore, the influence of PG on certain user experience metrics and interactive behaviors appears to be modulated by the recommendation domain, as evidenced by the interaction effects between the two examined factors.
This work contributes to the user-centered evaluation of ChatGPT-based CRS by investigating two prominent factors and offers practical design guidance. 

\end{abstract}




\settopmatter{printfolios=true}
\maketitle

\input{ch1_intro}

\input{ch2_related_work}

\input{ch3_experimental_setup}

\input{ch4_results_analysis}

\input{ch5_discussion}

\input{ch7_conclusion}
\bibliographystyle{ACM-Reference-Format}
\bibliography{references}

\appendix
\section{Post-study questionnaire}
\label{appendix}
Table~\ref{tbl:post-study-questionnaire} presents the question items in the post-study questionnaire, which is based on the short version of \textit{CRS-Que}~\cite{jin2024crs}.

\begin{table}[h]
    \caption{Post-study Questionnaire for Measuring the UX of CRS}
    \label{tbl:post-study-questionnaire}
    \footnotesize
    \begin{tabular}{p{70pt}p{150pt}}
    \toprule
    \textbf{Code} & \textbf{Description} \\
    \midrule
    \textbf{Perceived Qualities} & \\
    \hspace{0.6em}Accuracy & The recommended items were well-chosen. \\
    \hspace{0.6em}Novelty & The chatbot provided me with surprising recommendations that helped me discover new items that I wouldn’t have found elsewhere. \\
    \hspace{0.6em}Interaction Adequacy & I found it easy to tell the chatbot what I like/dislike. \\
    \hspace{0.6em}Explainability & The chatbot explained why the items were recommended to me. \\
    \hspace{0.6em}CUI Adaptability & I felt I was in sync with the chatbot. \\
    \hspace{0.6em}CUI Understanding & I found that the chatbot understood what I wanted. \\
    \hspace{0.6em}CUI Response Quality & Most of the chatbot’s responses make sense. \\
    \hspace{0.6em}CUI Attentiveness & The chatbot paid attention to what I was saying. \\
    \hline
    \textbf{User Beliefs} & \\
    \hspace{0.6em}Perceived Ease of Use & It was easy to find what I liked by using the chatbot. \\
    \hspace{0.6em}Perceived Usefulness & The chatbot gave me good suggestions. \\
    \hspace{0.6em}User Control & I felt in control of modifying my taste using this chatbot. \\
    \hspace{0.6em}Transparency & I understood how well the recommendations matched my preferences. \\
    \hspace{0.6em}CUI Humanness & The chatbot behaved like a human. \\
    \hspace{0.6em}CUI Rapport & The chatbot cared about me. \\
    \hline
    \textbf{User Attitudes} & \\
    \hspace{0.6em}Trust \& Confidence & I feel I could count on the chatbot to help me choose the items I need. \\
    \hspace{0.6em}Satisfaction & These recommendations made by the chatbot made me satisfied. \\
    \hline
    \textbf{Behavioural Intentions} & \\
    \hspace{0.6em}Intention to Use & I will use this chatbot frequently. \\
    \hspace{0.6em}Intention to try the recommendations & I would be likely to try the items recommended by the chatbot in the near future. \\
    \bottomrule
    \end{tabular}
\end{table}

\end{document}

%% file: ch1_intro.tex
\section{Introduction}

The emergence of advanced conversational AI like ChatGPT, which is based on large language models (LLMs) such as GPT-4, has 
reshaped user interaction with AI systems and significantly advanced the development of various real-world AI systems~\cite{nazir2023comprehensive}. As a typical AI system, recommender systems could benefit from enhancements through ChatGPT, as it facilitates a more natural language interaction with recommender systems and expands the scope for users to request recommendations spanning diverse domains~\cite{gao2023chat, spurlock2024chatgpt, liu2023chatgpt}. Some researchers have examined the capabilities and limitations of ChatGPT as a good recommender system~\cite{dai2023uncovering,liu2023chatgpt} and evaluated the ChatGPT-based recommendations regarding various aspects of recommendation quality, such as fairness, accuracy, diversity, stability, and temporal freshness~\cite{zhang2023chatgpt,deldjoo2024understanding}. 
Despite some limitations of ChatGPT in making recommendations, such as popularity bias, fairness, and authenticity, it exhibits significant promise in improving human interaction and recommendation accuracy.

The promise of ChatGPT in enhancing recommendation tasks has been a driving force behind the design and development of conversational recommender systems (CRS). CRS offer a stark contrast to the traditional recommender systems by enabling users to engage in multi-turn dialogues to refine their intentions and make decisions that align with their current preferences~\cite{jannach2021survey,gao2021advances}. The user experience (UX) of CRS is shaped not only by the system's ability to make recommendations but also by its ability to converse with the users, such as adaptability, understanding, and attentiveness~\cite{jin2021key,jin2024crs}. 
Recent research on ChatGPT-powered recommender systems has concentrated on evaluating the precision and relevance of the recommendations provided and some specific aspects of recommendation qualities, such as fairness and bias~\cite{deldjoo2024understanding, spurlock2024chatgpt, li2023preliminary}. 
However, a comprehensive understanding of the overall UX in CRS, as influenced by the integration of ChatGPT, remains under-explored. 
Evaluating recommender systems from the user perspective is crucial since user-perceived qualities of recommender systems can influence user attitudes (e.g., trust and satisfaction) and, ultimately, their behavioral intentions (e.g., intention to use and intention to purchase)~\cite{pu2011user}. Therefore, this work seeks to assess the UX of ChatGPT-based CRS through \textit{CRS-Que}~\cite{jin2024crs}, a user-centric evaluation framework for CRS. 

Moreover, this study investigates two significant factors that may impact the user experience of ChatGPT-based CRS, which are \textit{prompt guidance} (PG) and \textit{recommendation domain} (RD).
Despite the strong capabilities of ChatGPT, the performance of ChatGPT for a certain task heavily depends on the prompts used to interact with ChatGPT~\cite{lee2023few}. Appropriate prompts directly guide the model to generate the desired output, enabling ChatGPT to more accurately understand the task and thus enhance the conversational experience~\cite{korzynski2023artificial}. 
However, researchers have found that creating effective prompts could be difficult for non-expert users, which may hinder them from completing tasks or even lead to new problems (such as believing that the agent will "comprehend" the given prompt rather than just seeing the prompt as a trigger for the LLM, which leads to prompts like "do not say ABC" causing surprise as the agent saying ABC verbatim.) when using ChatGPT~\cite{zamfirescu2023johnny}. 
Thus, guiding users to compose effective prompts could facilitate the use of ChatGPT for fulfilling certain tasks like making recommendations~\cite{spurlock2024, gao:2023}. 
Another factor examined in this study is the recommendation domain. We concretely assessed two distinct recommendation domains differentiated by the level of decision-making stakes involved, as it was found in the previous work that they may affect users' perceptions of and interactions with the CRS~\cite{chen2013human, jameson2015human, elahi2022developing}. For high-stake decisions like seeking jobs or real estate advice~\cite{yuan2013toward,gutierrez2019explaining}, users generally aim to acquire comprehensive information and proceed with cautions. 
Conversely, users are inclined to decide swiftly with less deliberation for low-stake decisions, such as choosing books or music~\cite{gomez2015netflix}. 

Additionally, individual differences, such as a user's familiarity with the CRS, personality traits, and domain-specific knowledge, may lead to varied interactions with the same system setting across different recommendation domains~\cite{yan2023influence,cai2022impacts,jin2018effects}.
Considering these factors and individual user characteristics, we have formulated three research questions: 

\textbf{RQ1}: How does \textit{prompt guidance} (PG) influence the user experience (UX) of the ChatGPT-based CRS?

\textbf{RQ2}: How does \textit{recommendation domain} (RD) influence the UX of the ChatGPT-based CRS?

\textbf{RQ3}: How do \textit{personal characteristics} moderate the effects of PG and RD on the UX of the ChatGPT-based CRS?

To explore the research questions posed, we executed an online study with 100 participants, adopting a mixed-methods strategy that incorporates a between-subjects design to examine the influence of \textit{Prompt Guidance} (PG) being either enabled or disabled alongside a within-subjects design to assess the impact of \textit{Recommendation Domain} (RD) (book versus job recommendations). 
The findings indicate that both PG and RD significantly affect various user experience (UX) dimensions of recommendations, such as novelty, explainability, transparency, and the conversational adaptability of the UX~\cite{jin2021key, jin2024crs}. 
Notably, the influences of PG on factors like accuracy, adaptability, and user autonomy appear to vary between the recommendation domains. 
Furthermore, user familiarity with the recommendation system (RS) could moderate the effects of PG on explainability, transparency, and perceived ease of use.

This study's contributions are multifaceted: (1) It presents an online user study that evaluates the UX of ChatGPT-based conversational recommender systems, offering insights into ChatGPT's overarching impact on UX within this context. (2) It delves into the role of prompt guidance, a key design element in ChatGPT applications, evaluating its effect on UX across two distinct recommendation domains involving different levels of decision-making stakes. (3) The research further informs design strategies for personalization within such systems by examining how individual user characteristics may influence the interplay between PG and RD on the user experience of CRS.

%% file: ch2_related_work.tex
\section{Related work}

\subsection{ChatGPT-based CRS}
In contrast to traditional recommender systems, conversational recommender systems (CRS) engage users in interactive, multi-turn dialogues to elicit preferences and offer recommendations~\cite{jannach2021survey}, thereby enabling a more dynamic and natural refinement of user choices. Large language models (LLMs) have the potential to significantly enhance the computational and algorithmic modules of such systems~\cite{xu2024prompting,zhao2024recommender}. In this context, ChatGPT has showcased its proficiency in performing recommendation tasks. Initial evaluations of ChatGPT as a general recommendation agent have yielded promising outcomes in the realm of explainable recommendations, though its performance in sequential and direct recommendation scenarios appears to be less satisfying~\cite{liu:2023}. Subsequent research has assessed ChatGPT's capabilities of generating recommendations based on user preferences and the preferences of similar users, ranking items, and tackling the cold-start problem~\cite{di2023evaluating}. Recognizing ChatGPT's limitations in this domain, one study has integrated ChatGPT with conventional information retrieval ranking mechanisms to amplify its recommendation performance~\cite{dai2023uncovering}.

Furthermore, investigations have been conducted into how ChatGPT might enhance certain attributes of recommender systems, such as explainability~\cite{silva2024leveraging,liu:2023}, fairness~\cite{deldjoo2024understanding,zhang2023chatgpt}, diversity~\cite{deldjoo2024understanding}, and choice overload~\cite{kim2023decisions}. 
For instance, studies have highlighted ChatGPT's capacity to craft personalized and effective explanations for recommended items~\cite{silva2024leveraging,liu:2023}. In response to concerns about inherent biases in ChatGPT's recommendations~\cite{deldjoo2024understanding}, researchers have sought to refine the relevance of its suggestions and reduce popularity bias by leveraging user feedback and prompting techniques~\cite{spurlock2024}. Additionally, a novel benchmark has been proposed to assess the fairness of recommendations produced by large language models~\cite{zhang2023chatgpt}.

Despite the burgeoning research on the application of ChatGPT in recommendation tasks, there remains a lack of studies focusing on the user experience (UX) evaluation of ChatGPT-based conversational recommender systems. This study aims to fill this gap by utilizing \textit{CRS-Que}~\cite{jin2024crs}, a user-centric evaluation framework, to assess the UX of ChatGPT-based CRS across two distinct recommendation tasks.

\subsection{User Experience of CRS}
In contrast to offline evaluations, which prioritize objective metrics such as precision, recall, and item coverage, user-centered evaluations of conversational recommender systems (CRS) prioritize assessing the system's user experience from the end-user's perspective. Given the nature of CRS, the user experience (UX) constructs are employed to evaluate the perceived quality of both recommendations and conversational interactions within the system. A user-centric UX evaluation framework of CRS encompasses an assessment of users' perceived qualities of the system (e.g., novelty and CUI adaptability), user beliefs (e.g., perceived ease of use), user attitudes (e.g., trust and satisfaction), and behavioral intentions (e.g., intention to use)~\cite{jin2024crs}. It is shown that certain attributes of CRS (e.g., perceived accuracy and interaction adequacy) are particularly critical as they can influence users' intentional behaviors toward the recommended items, such as the intention to use or purchase them~\cite{jin2021key}.

Furthermore, personal characteristics are recognized as potentially significant moderating factors that can influence the impact of system design elements, such as algorithms and user interfaces, on the user experience of the recommender system~\cite{knijnenburg2012explaining}. Personalized experience design must take into account individual characteristics, including the user's experience with the system, domain knowledge, and personality traits, which may affect their perception of user control~\cite{jin2018effects,knijnenburg2011each}, the effectiveness of recommendation explanations~\cite{millecamp2019explain,chatti2022more}, and the level of user trust in the system~\cite{cai2022impacts}. Consequently, this study also considers the potential impacts of personal characteristics to evaluate the user experience of the ChatGPT-based CRS comprehensively.

\subsection{Prompt Guidance and Recommendation Domains}
Integrating Artificial Intelligence (AI) systems into various domains necessitates a user-centric approach to ensure effective and satisfactory interactions~\cite{shneiderman2022human}. This is particularly relevant in LLM-powered systems, where user prompts can shape the output of the system~\cite{henrickson2023prompting,liu2023pre}. For example, effective prompts for providing feedback could help users refine recommendations and mitigate bias~\cite{spurlock2024}. However, most non-technical users cannot effectively understand or control an AI system; thus, providing guidance could improve users' understanding and control over AI-driven processes~\cite{amershi2019guidelines}.
Likewise, non-technical users also find it difficult to compose effective prompts~\cite{zamfirescu2023johnny}. The knowledge of prompt engineering has been considered a new digital competence~\cite{korzynski2023artificial}. Therefore, prompt guidance or assistance could help users better articulate their needs, leading to more accurate and relevant generated results~\cite{khurana2024and}. In the realm of conversational agents, Myers et al.~\cite{myers2018patterns} demonstrate that guidance can improve user satisfaction by reducing the cognitive load required to interact with AI systems. Guidance in using AI systems, particularly in the context of prompt engineering, has emerged as a critical factor in enhancing user experience and system usability~\cite{ekin2023prompt}. Therefore, we hypothesize that \textit{prompt guidance} (PG) could influence the user experience of the ChatGPT-based CRS.

Moreover, the recommendation domain profoundly shapes user needs and preferences, which in turn guides UX design for recommender systems. Knijnenburg et al.~\cite{knijnenburg2012explaining} have discussed the importance of understanding the target audience's domain-specific requirements to tailor recommendation algorithms and interfaces accordingly. For instance, in the entertainment domain, users may prefer serendipitous recommendations to discover new content, while in e-commerce, the accuracy and trustworthiness of recommendations are critical~\cite{zhang2012auralist}. The interaction with recommender systems is also domain-specific. In travel and tourism, users often seek exploratory interfaces that allow them to browse and filter through recommendations~\cite{ricci2010mobile}. Conversely, in online retail, users may prefer a more trust-inspiring explanation interface that helps them to make purchase decisions quickly~\cite{pu2007trust}. In this study, we hypothesize that the \textit{recommendation domain} (RD) may influence the nature and extent of guidance required in the system. For instance, in domains with high complexity or high stakes, such as job recommendations, prompt guidance is not just a facilitator but a necessity for high-quality decision-making. 

%% file: ch3_experimental_setup.tex
\section{Experimental setup}
\label{experimental_setup}

Our study utilized ChatGPT as the recommendation agent, considering its capabilities in generating recommendations based on large language models~\cite{liu:2023,di2023evaluating}. 
The study aims to investigate how \textit{prompt guidance} influences the user experience of ChatGPT-based CRS in two different \textit{recommendation domains}. 
Therefore, we designed a 2 (withPG vs. withoutPG) x 2 (book recommendations vs. job recommendations) online user study to address our research questions. 

\subsection{Procedure}
\label{experimental_procedure}
Before starting the experiment, participants read the user study instructions to ensure that they understand the requirements of the study. After that, the participants are asked to fill out the pre-study questionnaire, which includes questions about their demographic properties and personal characteristics (see Section~\ref{questionnaires-pre}). Subsequently, the participants are randomly assigned to two experimental groups for prompt guidance (with or without). Each group is asked to use the assigned system to finish two recommendation tasks in different domains (book and job recommendations) (i.e., ``\textit{to find five items that suit your needs and add them to the wish list}). 
The order of the two tasks is randomized among all users to counter the learning effects in the within-subjects design. 
At the end of each task, participants are asked to fill out a post-study questionnaire based on the short version of \textit{CRS-Que}~\cite{jin2024crs}, a user-centric evaluation framework for conversational recommender systems.

\begin{figure*}[h]
    \centering
\includegraphics[width=.95\linewidth]{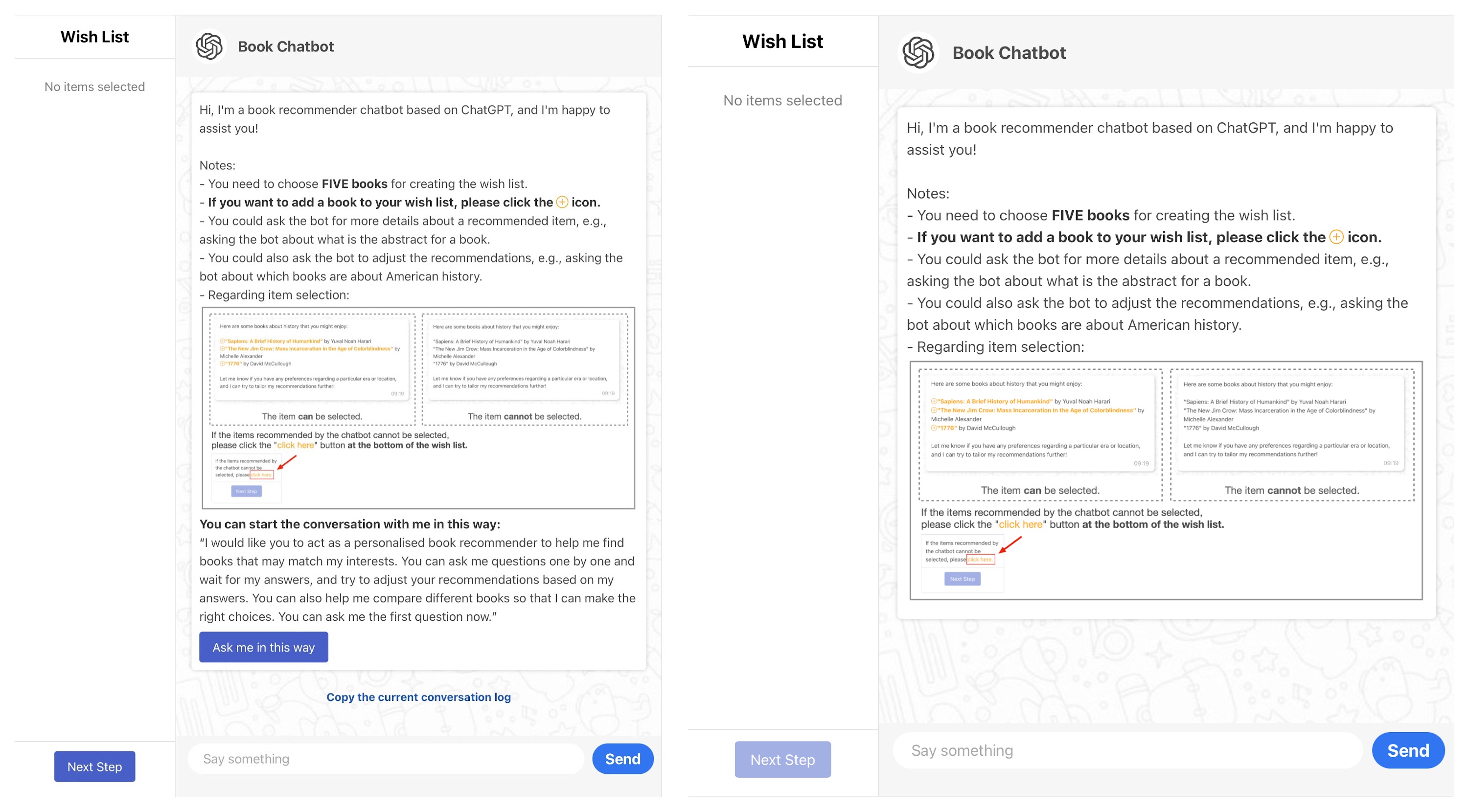}
    \caption{The ChatGPT-based CRS with prompt guidance (left) and without prompt guidance (right) for book recommendations.}
    \label{fig:chatbot_interface}
\end{figure*}

\subsection{Participants}
\begin{table}
\small
  \caption{Demographic Statistics of Participants}
  \label{tbl:participants}
  \begin{tabular}{lc}
    \toprule
        Property & Value \\
    \midrule
        \textbf{Age} & \\
        \hspace{0.6em}18 to 24 & 11 \\
        \hspace{0.6em}25 to 34 & 30 \\
        \hspace{0.6em}35 to 54 & 47 \\
        \hspace{0.6em}55 or over & 12 \\
        \hline
        \textbf{Gender} & \\
        \hspace{0.6em}Male & 47\\
        \hspace{0.6em}Female & 53\\
        \hline
        \textbf{Education} & \\
        \hspace{0.6em}Ph.D. & 1\\
        \hspace{0.6em}Master's degree & 15\\
        \hspace{0.6em}Completed some postgraduate & 2\\
        \hspace{0.6em}Bachelor's degree & 52\\
        \hspace{0.6em}Others & 30\\
  \bottomrule
\end{tabular}
\end{table}

We recruited participants from Prolific,\footnote{\url{https://www.prolific.com/}} a popular crowdsourcing platform for academic research. As the experimental system's primary language is English, we recruited 100 participants from English-speaking countries such as the UK and the USA. To ensure the quality of participants, our recruitment targets participants who have finished at least 100 tasks and the approval rate should be greater than 95\%. The experiment took around 20 minutes on average. We provided 1.5 British Pounds to the participant who successfully completed. 
The research ethics committee at the authors' university approved this experiment. 
Table~\ref{tbl:participants} shows the demographic information of all participants. Upon reviewing the answers to attention-check questions, as well as the duration and substance of the conversations, we have determined that all participants adhered to the study guidelines, successfully completing the study.

\subsection{Design Manipulations}
\label{ivs}
We manipulated the prompt's initial setting and the recommendation domain 
to investigate their impacts on the user experience of the system.

\subsubsection{Prompt Guidance (PG)}
In our study, PG refers to an example of crafting queries that clearly specify the context, constraints, and preferences for the recommendations users may seek. We designed prompt guidance according to the key elements of an effective prompt~\cite{ekin2023prompt,henrickson2023prompting}, such as showing clear intention, including specific details, providing a context where necessary, and the prompt engineering guide provided by OpenAI.\footnote{\url{https://platform.openai.com/docs/guides/prompt-engineering}} This study examines two conditions in PG:\\
\begin{itemize}
    \item \textbf{withoutPG}: The system does not display a prompt example, and participants use their own words to ask the system to make recommendations.   For example, to get book recommendations, participants may ask \textit{"Can you recommend some fun new fiction books for me? I like books of suspense."}
    \item \textbf{withPG}: The system provides a pre-designed prompt at the start of the conversation. It depicts the task details and requires the system to act as a personalized recommender: \textit{"I would like you to act as a personalized xxx recommender to help me find xxx that may match my interests. You can ask me questions one by one and wait for my answers, and try to adjust your recommendations based on my answers. You can also help me compare different xxx so that I can make the right choices. You can ask me the first question now."} (where xxx is \textit{book} or \textit{job}, depending on the assigned recommendation domain). The offered prompt example could enhance the quality of recommendations and conversations by clarifying preferences, providing context, and encouraging detailed responses. 
\end{itemize}

\subsubsection{Recommendation Domain (RD)}
We determine two distinct recommendation domains (i.e., \textbf{book recommendations} and \textbf{job recommendations}) and have explored how they affect the user experience, particularly focusing on the perceived stakes involved in the recommendation. By comparing these two domains, the study can reveal how the system design (e.g., prompt guidance) needs to adapt to different stakes levels. For low-stake recommendations like books, the system might prioritize discovery, diversity, and personalization with less emphasis on detailed information~\cite{li2023bookgpt}. In contrast, the system may need to provide more comprehensive, detailed, and carefully vetted information for high-stake recommendations, ensuring users feel confident and secure in their decision-making process~\cite{gutierrez2019explaining}. Participants are instructed to converse with the ChatGPT-based CRS for each domain to obtain recommendations and add five items to their wish list. We keep the system setting and the study task consistent for two recommendation domains to investigate how RD interacts withPG to influence the user experience of the ChatGPT-based CRS. 

\subsection{User Interfaces}
Figure~\ref{fig:chatbot_interface} shows the user interfaces of the ChatGPT-based CRS for book recommendations. The user interface contains two panels. The left panel is the wish list that shows the items the user selects and the ratings of selected items. The right panel refers to the conversation window, where participants follow the task instructions to perform the recommendation task by conversing with the system. 
In the condition with prompt guidance, the pre-defined prompt will appear below the welcome message (as highlighted in the red box in Figure~\ref{fig:chatbot_interface} [left]); by clicking the ``\textit{Ask me in this way}'' button, the suggested prompts will be copied to the input box for participants to use. In contrast, the version without prompt guidance requires participants to compose their own prompts. For example, the participant may ask ``\textit{Please recommend some books on European history to me.}'' During the conversation, they need to answer questions generated by ChatGPT to provide their preferences. They are encouraged to request that ChatGPT provide more details about the recommended results (such as an explanation of the item) or to provide feedback. The recommended item (e.g., a book or job name) is highlighted in orange with an icon $\oplus$ placed next to it so that the participants can add it to their wish list (Figure~\ref{fig:chatbot_interface} [right]). Participants then need to select five favorite items from the recommended results provided by ChatGPT and add them to their wish list, after which they can click the "Next Step" button to finish the conversation.


\subsection{Measurements}
\subsubsection{Pre-study Questionnaire}
\label{questionnaires-pre}
The pre-study questionnaire contains questions to collect participants' demographic information and eight questions that measure four personal characteristics (PC), including \textit{experience with RS}, \textit{trust propensity}, \textit{tech-savviness}, and \textit{experience with ChatGPT} (see Table \ref{tbl:pre-study-questionnaire}). Most of these questions have been used in previous user studies on the personalization of recommender systems. For example, users with more experience may navigate the system more efficiently and understand the nuances of the recommendations better, leading to a more positive user experience~\cite{knijnenburg2012explaining,konstan2012recommender}. Users with higher trust propensity may be more likely to try out recommendations and rate the system more favorably~\cite{knijnenburg2011each,cai2022impacts}. Each question is rated on a 5-point Likert scale from 1 (strongly disagree) to 5 (strongly agree). 


\begin{table}
\small
\caption{The Pre-study Questionnaire}
\label{tbl:pre-study-questionnaire}
\begin{tabular}{p{70pt}p{140pt}}
\toprule
  PC & Statement  \\ 
\midrule
    Experience with RS & Q1. \textit{I am familiar with recommender systems (e.g., book recommendations from Amazon or job recommendations from LinkedIn).} \\
    & Q2. \textit{I have used recommender systems frequently.} \\
    
    \hline
    
    Trust propensity & Q3. \textit{I tend to trust a new technology, even though I have little knowledge of it.} \\
    & Q4. \textit{Trusting someone or something is difficult.} \\
    
    \hline
      
    Tech-savviness & Q5. \textit{I am confident when it comes to try a new technology.} \\
    
    \hline 
    Experience with ChatGPT & Q6. \textit{I consider myself to be an expert in using ChatGPT.}\\
    & Q7. \textit{I am knowledgeable about ChatGPT.} \\
    & Q8. \textit{I have extensive experience in using ChatGPT.} \\
\bottomrule
\end{tabular}
\end{table}


\subsubsection{Post-study Questionnaire}
\label{questionnaires-post}
To assess the user experience in this experiment, we adopted the short version of the CRS-Que framework with 18 questions for measuring the UX of conversational recommender systems (see Appendix~\ref{appendix})~\cite{jin2024crs}. The items with a CUI prefix (Conversational User Interface) indicate the metrics for measuring user experience for conversations in \textit{CRS-Que}. This framework assesses the perceived qualities of both recommendations and conversations (e.g., accuracy, novelty, CUI understanding, CUI adaptability) of a CRS. Furthermore, it investigates how these qualities could influence higher-level UX dimensions: User Beliefs (e.g., ease of use, CUI humanness, and CUI rapport), User Attitudes (e.g., trust and satisfaction), and Behavioral Intentions (e.g., intention to use). As with a pre-study questionnaire, all questions are rated on a 5-point Likert scale.

\subsubsection{Interaction Logs}
When using ChatGPT for recommendations, interaction logs are helpful for understanding user behaviors and explaining the system's user experience.
Therefore, we captured users' conversational interaction behavior by recording the number of dialog turns in each conversation, the total number of words, the conversation duration, and the average number of words per dialog turn (see Table \ref{tbl:interaction-data-analysis-table}).

%% file: ch4_results_analysis.tex
\section{Results}\label{}

\begin{figure}[H]
    \centering
    \includegraphics[width=.9\linewidth]{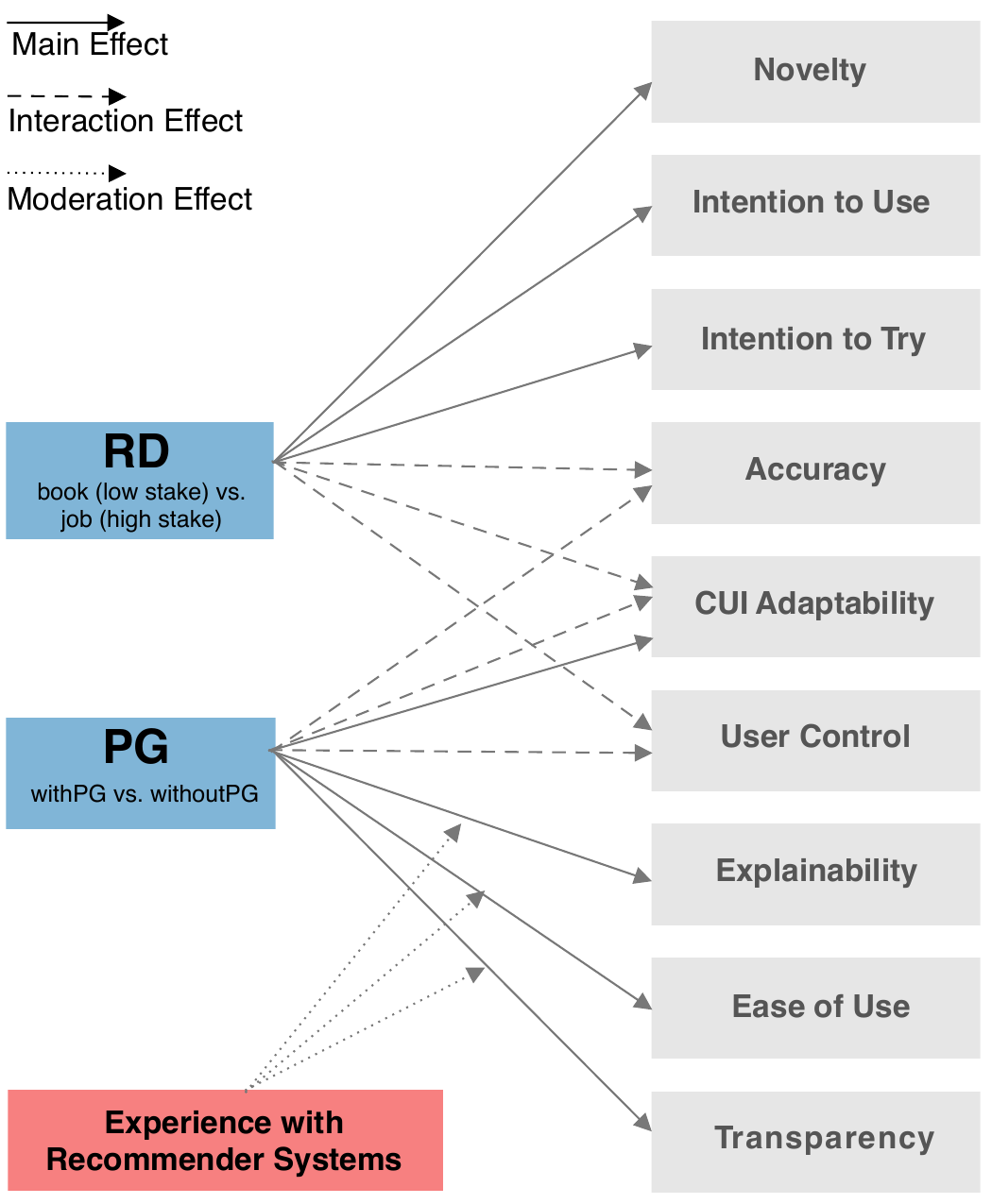}
    \caption{The summary of effects of prompt guidance (PG) and recommendation domain (RD)}
    \label{fig:effects_graph}
\end{figure}

\begin{figure*}[ht]
    \centering
    \includegraphics[width=1\linewidth]{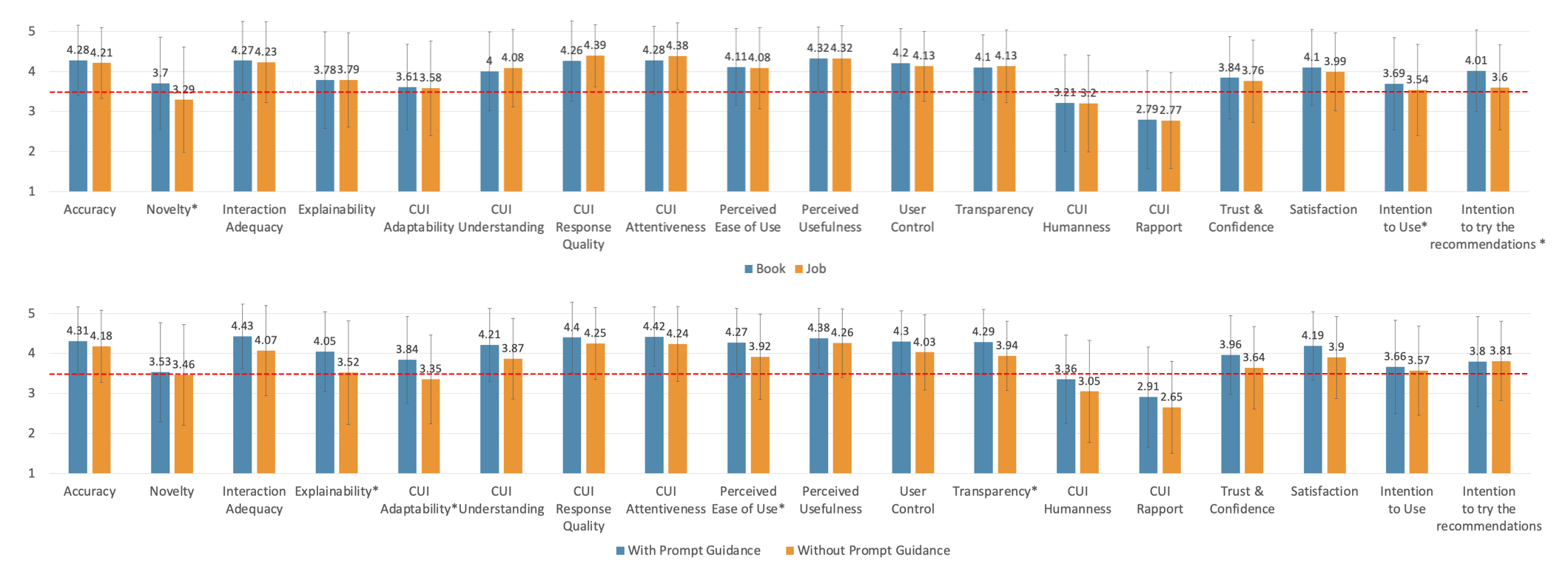}
    \caption{Results of the UX measurements of CRS grouped by the conditions of two independent variables: RD and PG. A cut-off value of 3.5 represents agreement on the five-point Likert scale. * is marked for significant difference at the 5\% level (p-value < 0.05).}
    \label{fig:post-study-questionnaire-chart}
\end{figure*}

\begin{figure*}[h]
    \centering
\includegraphics[width=.9\linewidth]{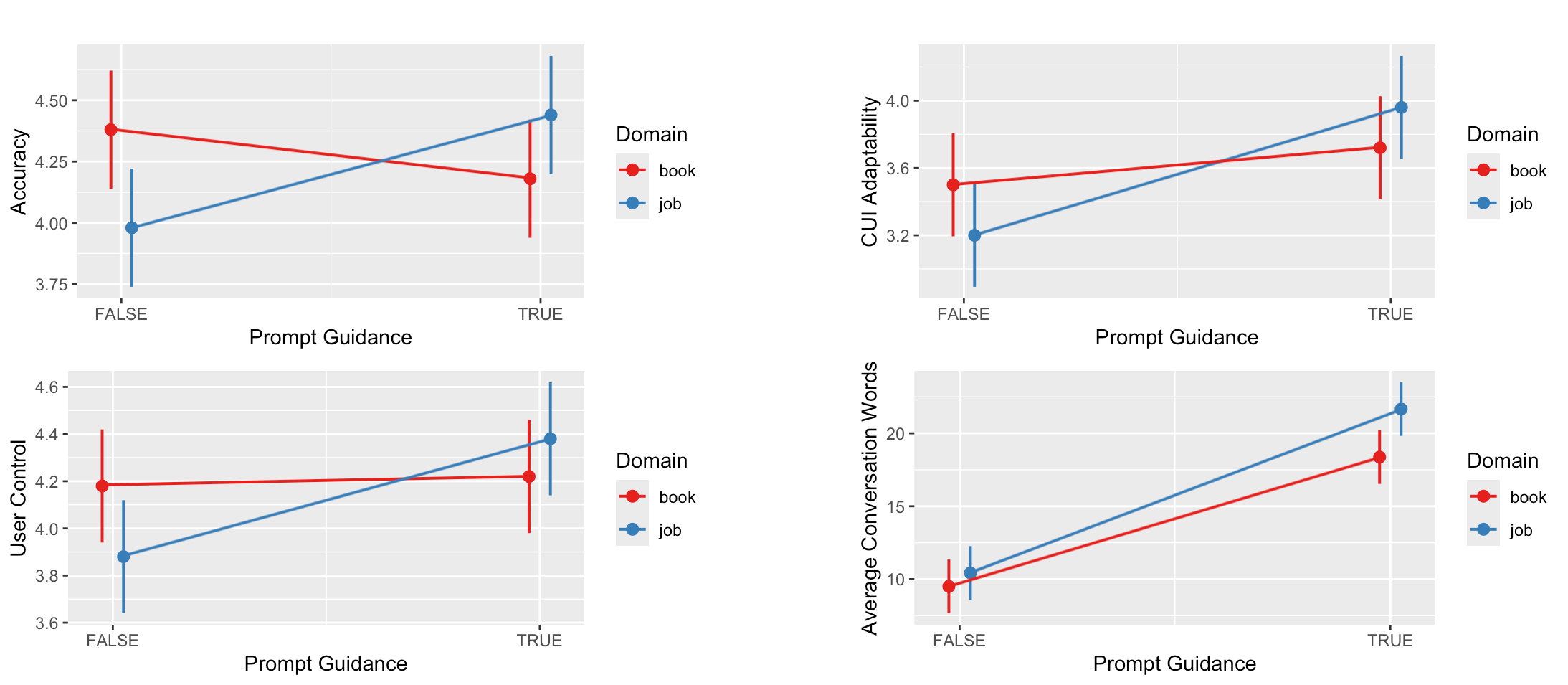}
    \caption{Interaction Effects of PG and RD on Accuracy (top-left),  User Control (bottom-left), CUI Adaptability (top-right), and the Average Words per Conversation (bottom-right).}
    \label{fig:interaction_effects_graphs}
\end{figure*}

This section presents findings derived from users' responses to the questionnaire to assess the overall user experience of ChatGPT-based CRS. Figure~\ref{fig:effects_graph} summarizes all the significant effects of two independent variables (i.e., PG and RD), including main effects and interaction effects, as well as the significant moderation effects of personal characteristics on the main effects of PG and RD. The descriptive statistics of the user experience measured by \textit{CRS-Que} is shown in Figure~\ref{fig:post-study-questionnaire-chart}. Table~\ref{tbl:interaction-data-analysis-table} presents the descriptive statistics of interaction behavior metrics.

\begin{table}
\small
\caption{Descriptive Statistics of Interaction Behavior Metrics}
\label{tbl:interaction-data-analysis-table}
    \begin{tabular}{lllp{38pt}p{38pt}}
    \toprule
         & \textbf{Book} & \textbf{Job} & \textbf{withPG} & \textbf{no PG} \\
    \midrule
        \textbf{Turns} & 6.51 & 4.66 & 6.44 & 4.73 \\
              \hline
        \textbf{Total Words} & 85.59 & 73.31 & 113.38 & 45.52 \\
              \hline
        \textbf{Duration} & 266.65 & 209.5 & 267.77 & 208.38 \\
              \hline
        \textbf{AVG Words} & 13.937 & 16.046 & 20.017 & 9.967 \\
    \bottomrule
    \end{tabular}
\end{table}

\subsection{Main Effects}
As the data do not conform to normal distribution according to the Shapiro–Wilk test result, we chose a non-parametric approach to factorial ANOVA, ART-rank, to analyze the interaction as well as the main effects of two independent variables (IVs)~\cite{wobbrock2011aligned}. 

\subsubsection{The Effects of Prompt Guidance (PG)} 

The results from the ART-rank test indicate that the PG has a significant impact on the perceived qualities and user beliefs, such as explainability, conversational user interface (CUI) adaptability, perceived ease of use, and transparency, at a significance level of 0.05 (Figure~\ref{fig:effects_graph}). Particularly, the explainability is significantly higher in the withPG group (Mean = 4.050, SD = 0.999) than in the withoutPG group (Mean = 3.520, SD = 1.299), $F(1, 98) = 6.316, p < .05, \eta^{2} = 0.050$. 
Users feel more sync in the withPG group (Mean = 3.840, SD = 1.089) than in withoutPG group (Mean = 3.350, SD = 1.114), $F(1, 98) = 0.009, p < .001, \eta^{2} = 0.048$. Additionally, the perceived ease of use is significantly higher in the withPG group (Mean = 4.270, SD = 0.863) than in the withoutPG group (Mean = 3.920, SD = 1.070), $F(1, 98) = 4.086, p < .05, \eta^{2} = 0.032$; and transparency is rated significantly higher in the withPG group (Mean = 4.290, SD = 0.820) than in the withoutPG group (Mean = 3.940, SD = 0.862), $F(1, 98) = 6.55, p < .05, \eta^{2} = 0.042$.

\subsubsection{The Effects of Recommendation Domain (RD)}
The test results show that the recommendation domain significantly influences perceived quality and behavioral intentions (i.e., novelty, intention to use, and intention to try the recommendations) at an alpha level of 0.05. Specifically, the perceived novelty is significantly higher in book recommendations (Mean = 3.7, SD = 1.15) than in job recommendations (Mean = 3.29, SD = 1.313), $F(1, 98) = 7.55, p < .01, \eta^{2} = 0.027$. In addition, participants tend to use the system for book recommendations (Mean = 3.69, SD = 1.152) than for job recommendations (Mean = 3.54, SD = 1.141), $F(1, 98) = 4.006, p < .05, \eta^{2} = 0.004$; and they tend to try the recommended books (Mean = 4.01, SD = 1.02) than the recommended jobs (Mean = 3.6, SD = 1.064), $F(1, 98) = 17.169, p < .001, \eta^{2} = 0.038$. The results show that both PG and RD do not significantly influence the metrics of interaction behavior.




\begin{figure*}[h]
    \centering
\includegraphics[width=.9\linewidth]{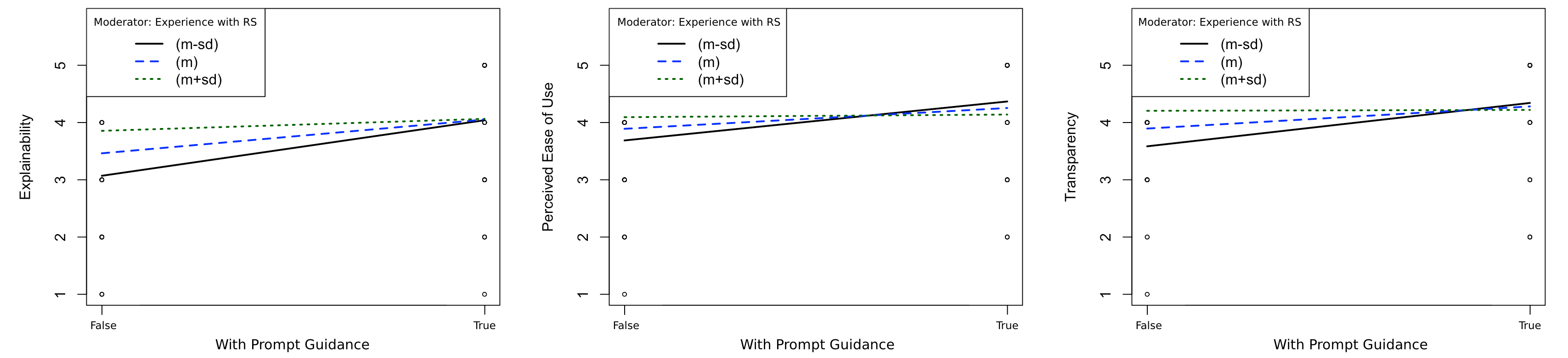}
    \caption{Moderation effects of the personal characteristic ``experience with recommender systems'' on UX aspects explainability (left), perceived ease of use (middle), and transparency (right).}
    \label{fig:moderation_effects_graphs}
\end{figure*}

\subsection{Interaction Effect}
In addition to the main effects, the ART-rank test reveals the combined effects of two independent variables (PG and RD) on the various user experience aspects of the system. The two associated dashed arrow lines in Figure~\ref{fig:effects_graph} indicate the significant interaction effects on Accuracy, CUI Adaptability, and User Control. More specifically, as shown in Figure~\ref{fig:interaction_effects_graphs} (top-left), PG allows participants to perceive higher accuracy in job recommendations, whereas an adverse effect is shown in book recommendations, $F(1, 98) = 8.725, p < .01, \eta^{2} = 0.036$. Similarly, PG allows participants to perceive higher control in job recommendations but negatively affects the perceived control in book recommendations (Figure~\ref{fig:interaction_effects_graphs} [bottom-left]), $F(1, 98) = 6.556, p < .05, \eta^{2} = 0.018$. However, for CUI adaptability, PG shows the opposite interaction effects with two domains: PG leads to higher adaptability in job recommendations but lower adaptability in book recommendations (Figure~\ref{fig:interaction_effects_graphs} [top-right]), $F(1, 98) = 6.000, p < .05, \eta^{2} = 0.014$. In addition, we find a significant interaction effect on interaction behavior: PG allows participants to express more for seeking job recommendations than for book recommendations, whereas the difference is not significant if PG is not provided (Figure~\ref{fig:interaction_effects_graphs} [bottom-right]), $F(1, 98) = 5.525, p < .05, \eta^{2} = 0.005$.

\subsection{Moderation Effect}
Furthermore, we perform a moderation analysis to see how our investigated personal characteristics could moderate the significant main effects of independent variables on user experience. The linear regression models reveal significant moderation effects of \textit{the experience with RS} on the effects of PG on three UX aspects: Explainability ($Estimate = -0.192, SE = 0.087, p < .05$), Perceived Ease of Use ($Estimate = -0.16, SE = 0.074, p < .05$), and Transparency ($Estimate = -0.187, SE = 0.063, p < .01$), as illustrated by the black dotted arrow lines in Figure~\ref{fig:effects_graph}. To better illustrate moderation effects, we use an R package, rockchalk~\footnote{\url{https://cran.r-project.org/web/packages/rockchalk/index.html}}, to plot the simple slopes (1 SD above and 1 SD below the mean) for analyzing the moderating effect. Figure~\ref{fig:moderation_effects_graphs} illustrates that the beneficial impacts of prompt guidance (PG) on perceived explainability (left), ease of use (middle), and transparency (right) are especially pronounced among participants who have less experience with recommendation systems (RS), as indicated by the black line.



%% file: ch5_discussion.tex
\section{Discussion}
Our findings indicate that prompt guidance (PG) significantly enhances the user experience of the ChatGPT-based CRS. Furthermore, the recommendation domain (RD) could influence participants' perception of and interaction with the system. The following section will discuss the key findings based on the main effects, interaction effects, and moderation effects, as well as the limitations of this study. 

\subsection{Key Findings}
\textbf{Key Finding \#1: The prompt guidance (PG) could make the system better align with user needs and help users grasp the rationale behind the recommendations, thereby simplifying the system's use (RQ1).}
This finding implies that ChatGPT-based CRS should provide PG rather than letting users compose the prompts, which aligns with other studies on prompt engineering. For example, prompt engineering involves embedding a description of the task participants want the system to perform into the input rather than relying on the system to decipher unclear expressions that participants might employ \cite{liu2023pre}. The value of prompt guidance lies in its function as a tool that aids participants in guiding ChatGPT to produce higher quality, more usable, and satisfactory answers; it instructs the system on how to search and generate responses within a vast information space \cite{henrickson2023prompting}. In Section~\ref{ivs}, we demonstrate how our designed prompt guidance requires ChatGPT to engage in step-by-step active questioning based on the recommendation task at hand, then adjust its responses according to user feedback, and facilitate decision-making by presenting comparisons between different items. This process clarifies the entire interaction, enabling participants to express their preferences under the guidance of ChatGPT and receive positive feedback from the system, leading to satisfactory decision-making. Conversely, in the absence of prompt guidance, participants were merely able to express their demands effectively as most non-technical people do not have sufficient knowledge in prompt engineering~\cite{khurana2024and,zamfirescu2023johnny}. 
Even though some participants could write good prompts for recommendation tasks, they may need to expend considerable effort testing different prompts to enable ChatGPT to understand their preferences accurately and provide recommendations. Crafting effective prompts is a challenging task even for expert users, and some studies highlight the critical role of prompts and reflect the difficulties encountered in the process of prompt creation~\cite{lee2023few}.

\textbf{Key Finding \#2: Users tend to perceive more novel recommendations and are inclined to use and try the system in a low-stake recommendation domain (e.g., book recommendations) (RQ2).} The finding suggests that ChatGPT or similar AI systems tend to help users explore novel recommendations in low-stake domains, such as book recommendations. In low-stake scenarios, users are generally more open to exploring new options because the cost of a less-than-ideal outcome is low. For instance, if a book recommended by ChatGPT doesn't meet a user's expectations, the consequence is relatively minor (time spent reading a few pages or chapters), encouraging users to try out new recommendations they might not have considered otherwise. In addition, users might value ChatGPT's ability to introduce them to items they haven't encountered before for low-stake recommendations. This perception can make interactions with the generative AI system more engaging, as users view the system as a tool for discovery and learning~\cite{filippi2023measuring,baabdullah2024generative}. In contrast, users tend to be more conservative in high-stake domains such as job recommendations. High-stake decisions carry significant consequences. Users need to trust the source of the recommendation when the stakes are high. The bias, fairness, and ethics issues in large language models may hinder users from trusting ChatGPT in high-stake domains, such as healthcare and education~\cite{choudhury2024large,kasneci2023chatgpt}.

\textbf{Key Finding \#3: The decision to provide PG may depend on the recommendation domain. (RQ1 \& RQ2).}
The interaction effects of PG and RD on the accuracy, user control, adaptability, and average words suggest that PG is particularly beneficial in high-stake recommendation tasks, which could be attributed to the nature of high-stake recommendation tasks involving complex decisions with multiple factors to consider~\cite{kunreuther2002high}. Given the costly implications of decisions in high-risk domains, participants would like to interact with the system more and heighten expectations for accuracy \cite{elahi2022developing}. PG ensures the quality of the prompt for recommendation tasks, enabling ChatGPT to refine user preferences and continually align the conversation with the updated preferences. This implicit adaptation to user preference could also increase a sense of user control. In contrast, the impact of PG is relatively mundane, showing no substantial improvement and even a decrease in accuracy, which may be attributed to some unknown constraints imposed by PG or the hallucination of ChatGPT that generates nonexistent books~\cite{buchanan2024chatgpt}, as pointed out by some participants. Therefore, the prompt guidance ought to be tailored to the stake level pertinent to recommendation domains, which demands consideration of differing objectives and feedback mechanisms for a specific recommendation task. Such considerations permit ChatGPT to fine-tune its interactive style suitably, addressing users' unique requirements.  

\textbf{Key Finding \#4: The tailored configuration of PG could better accommodate users who are familiar with recommender systems to varying degrees (RQ3).}
The moderating effects of the experience with RS indicate that prompt guidance is especially beneficial for novices, enhancing their perception of the system's explainability, transparency, and user-friendliness. Moreover, PG allows the system to adapt conversations based on the user's RS experience level. For well-versed users in RS, PG might streamline the process, offering less guidance and allowing for more user autonomy. Conversely, for those less familiar, it could provide more comprehensive guidance and educational cues~\cite{bell2002adaptive}.



\subsection{Limitations}
This study contains three major limitations. \textit{First}, our primary participants recruited from the Prolific platform are all native English speakers. Therefore, this experiment may not be able to reflect the user experience of such a system from a non-native English speaker's perspective. Non-native English speakers may benefit more from prompt guidance. \textit{Second}, the analysis of results mainly focuses on quantitative data. Future analysis will associate with the qualitative data to have a deeper understanding of user behavior and perceptions of such a system. \textit{Third}, despite referencing a large amount of relevant work in the design process of prompt guidance and conducting extensive testing and optimization before the experiment, they may still not enable ChatGPT to perform at its best in recommendation tasks.

%% file: ch7_conclusion.tex
\section{Conclusion}

This paper primarily investigates the impact of PG and RD on the overall user experience of a system. Utilizing a mixed-method online empirical study, where participants were exposed to a CRS with varying combinations of these factors, it was discovered that both PG and RD could significantly influence several user experience aspects of CRS. The interaction effects and moderation effects suggest the presence of PG should consider the recommendation domain and the user's experience level of recommender systems. Our study substantiates the significant roles of PG and RD in shaping the user experience in ChatGPT-based CRS. In the realm of prompt engineering for these systems, it is essential to consider the distinct expectations and behaviors users exhibit across various application domains, as well as to consider users' context in a comprehensive design approach. This work contributes to the user-centered evaluation of ChatGPT-based CRS by exploring two influential factors and provides insights into the user experience design of ChatGPT-based CRS.